\documentclass[letter]{aa}
\usepackage[varg]{txfonts}
\usepackage{graphicx}
\usepackage{natbib,twoopt}
\usepackage[breaklinks=true]{hyperref} 
\usepackage{lscape}
\bibpunct{(}{)}{;}{a}{}{,} 
\makeatletter
 \newcommandtwoopt{\citeads}[3][][]{\href{https://ui.adsabs.harvard.edu/abs/#3/abstract}%
 {\def\hyper@linkstart##1##2{}%
 \let\hyper@linkend\@empty\citealp[#1][#2]{#3}}}
 \newcommandtwoopt{\citepads}[3][][]{\href{https://ui.adsabs.harvard.edu/abs/#3/abstract}%
 {\def\hyper@linkstart##1##2{}%
 \let\hyper@linkend\@empty\citep[#1][#2]{#3}}}
 \newcommandtwoopt{\citetads}[3][][]{\href{https://ui.adsabs.harvard.edu/abs/#3/abstract}%
 {\def\hyper@linkstart##1##2{}%
 \let\hyper@linkend\@empty\citet[#1][#2]{#3}}}
 \newcommandtwoopt{\citeyearads}[3][][]%
 {\href{https://ui.adsabs.harvard.edu/abs/#3/abstract}
 {\def\hyper@linkstart##1##2{}%
 \let\hyper@linkend\@empty\citeyear[#1][#2]{#3}}}
\makeatother

\begin{document}

   \title{Basaltic quasi-mini-moon: Characterizing 2024~PT$_{5}$ with the 10.4~m Gran Telescopio 
          Canarias and the Two-meter Twin Telescope\thanks{Based on observations made with the 
          Gran Telescopio Canarias (GTC) telescope, in the Spanish Observatorio del Roque de los 
          Muchachos of the Instituto de Astrof\'{\i}sica de Canarias (IAC, program IDs 19-GTC16/24B 
          and 63-GTC56/24B), and the Two-meter Twin Telescope (TTT), in the Spanish Observatorio del 
          Teide of the IAC (PEI project SIDERA24).}
         }
   \author{R.~de~la~Fuente Marcos\inst{1}
            \and
           J.~de~Le\'on\inst{2,3}
            \and
           M. Serra-Ricart\inst{2,3,4}
            \and
           C.~de~la~Fuente Marcos\inst{5}
            \and
           M.~R. Alarcon\inst{2,3}
            \and
           J. Licandro\inst{2,3}
            \and
           S. Geier\inst{6,2}
            \and
           A. Tejero\inst{6,2}
            \and
           A. Perez Romero\inst{6,2}
            \and
           F. Perez-Toledo\inst{6,2}
            \and
           A. Cabrera-Lavers\inst{6,2,3}
          }
   \authorrunning{R. de la Fuente Marcos et al.}
   \titlerunning{Characterization of 2024~PT$_{5}$ with GTC and TTT} 
   \offprints{R. de la Fuente Marcos, \email{rauldelafuentemarcos@ucm.es}}
   \institute{AEGORA Research Group,
              Facultad de Ciencias Matem\'aticas,
              Universidad Complutense de Madrid,
              Ciudad Universitaria, E-28040 Madrid, Spain
              \and
              Instituto de Astrof\'{\i}sica de Canarias (IAC),
              C/ V\'{\i}a L\'actea s/n, E-38205 La Laguna, Tenerife, Spain
              \and
              Departamento de Astrof\'{\i}sica, Universidad de La Laguna,
              E-38206 La Laguna, Tenerife, Spain
              \and
              Light Bridges S.L.,
              Observatorio del Teide, Carretera del Observatorio s/n, E-38500 Guimar, Tenerife, Canarias, Spain
              \and
              Universidad Complutense de Madrid,
              Ciudad Universitaria, E-28040 Madrid, Spain
              \and
              GRANTECAN,
              Cuesta de San Jos\'e s/n, E-38712 Bre\~na Baja, La Palma, Spain
             }
   \date{Received 23 October 2024 / Accepted 20 January 2025}
   \abstract
      {Small bodies in Earth-like orbits, the Arjunas, are good targets for 
       scientific exploration, and space mining or in situ resource 
       utilization (ISRU) trials as they enable low-cost missions. The subset 
       of these objects that experience recurrent temporarily captured flyby 
       or orbiter episodes are among the best ranked in terms of 
       accessibility. Only a handful of objects are known to have engaged in 
       such a dynamical behavior. Finding and characterizing more of them may 
       help to expand scientific and commercial research activities in space 
       over the next few decades. Asteroid 2024~PT$_{5}$ is a recent finding 
       that shows dynamical traits in common with this group.  
       }
      {Here we investigate the orbital context of 2024~PT$_{5}$ and its 
       spectral and rotational properties. 
       }
      {We studied the short-term orbital evolution of 2024~PT$_{5}$ using 
       direct $N$-body simulations. We identified its spectral class from the 
       visible reflectance spectrum and used photometric observations to
       derive its rotational properties. Observational data were obtained 
       with the OSIRIS camera spectrograph at the 10.4~m Gran Telescopio 
       Canarias and the Two-meter Twin Telescope. 
       }
      {Asteroid 2024~PT$_{5}$ experiences recurrent co-orbital engagements 
       and episodes in which it has negative geocentric orbital energy while 
       inside a geocentric distance under three Hill radii, which we call 
       quasi-mini-moon events. Its visible spectrum is consistent with that of 
       lunar-like silicates. Photometric data suggest a rotation period 
       $\lesssim$1~h.
       }
      {The discovery of 2024~PT$_{5}$ confirms that events resembling 
       temporary captures are relatively frequent and involve objects larger 
       than a few meters, suitable as accessible targets for scientific 
       research activities and demonstrating ISRU technologies. 
       }


   \maketitle

   \section{Introduction\label{Intro}}
      Both ESA and NASA are pushing into the commercial space sector supporting a nascent global space economy (see, e.g., 
      \citealt{2023PNAS..12022013R,2024AcAau.222..162P}). Scientific exploration and mining or in situ resource utilization (ISRU) of 
      asteroids are activities well suited for development within the context of this fast-evolving sector (see, e.g., 
      \citealt{2013aste.book..151G,SERCEL2018477,2021AcAau.181..249X}), but only if there are enough targets for low-cost missions. 
      NASA's Near-Earth Object Human Space Flight Accessible Targets Study (NHATS; 
      \citealt{2012LPI....43.2842A})\footnote{\href{https://cneos.jpl.nasa.gov/nhats/}{https://cneos.jpl.nasa.gov/nhats/}} is compiling a 
      list of near-Earth objects (NEOs) that identifies potential targets for future exploration, but such a database must include 
      information on the composition of the targets for better mission planning. Reflectance spectroscopy can help in finding what these 
      more accessible targets are made of \citep{2020EPSC...14..203L,2021plde.confE..25D}. 

      NHATS includes thousands of targets, but few have been characterized spectroscopically \citep{2019A&A...627A.124P}. Most targets with 
      the highest number of viable trajectories (the most accessible ones) follow heliocentric orbits with periods close to one sidereal 
      year and have low eccentricity and low inclination. They are part of a secondary asteroid belt that surrounds the orbit of the 
      Earth--Moon system and define a dynamical class, the Arjunas \citep{1993Natur.363..704R}. Within the 25 most accessible targets, a 
      handful have recurring temporarily captured flyby or orbiter events, also called mini-moon episodes \citep{2012Icar..218..262G}.  

      The recently found asteroid 2024~PT$_{5}$ is part of the group of very accessible targets \citep{2024RNAAS...8..224D}. Here we used 
      reflectance spectroscopy, photometry, and $N$-body simulations to confirm its nature and dynamical context. This Letter is organized 
      as follows. In Sect.~\ref{Data} we introduce the background of our research, and present the data and tools used in our analyses. In 
      Sect.~\ref{Results} we investigate whether 2024~PT$_{5}$ is natural or artificial and its probable origin. In Sect.~\ref{Discussion} 
      we discuss our results.  Section~\ref{Conclusions} summarizes our conclusions. The Appendices include supporting material.

   \section{Context, methods, and data\label{Data}}
      In this section we revisit the dynamical concepts that are used in our analysis. Software tools and data are also discussed here. 

      \subsection{Dynamics background}
         For this work (but see Appendix~\ref{HillR}) we adopted the criteria in \citet{1996Icar..121..207K} and \citet{2012Icar..218..262G}. 
         Temporarily captured natural irregular satellites of Earth must have negative geocentric energy, and their geocentric distance must 
         be under three Hill radii (Earth's Hill radius is $\sim$0.01~au); in other words, the intruding NEO must approach at close range 
         ($<$0.03~au) and low relative velocity ($\lesssim$1~km~s$^{-1}$). The only NEOs that can regularly meet the capture conditions 
         defined by \citet{1996Icar..121..207K} are the Arjunas. Some of them also experience resonant behavior induced by the 1:1 
         mean-motion resonance (see, e.g., \citealt{2013MNRAS.434L...1D}) in which their relative mean longitude with respect to our planet 
         ($\lambda_{\rm r}$) oscillates about a fixed value \citep{2002Icar..160....1M} to become Earth co-orbitals as their orbital periods 
         closely match that of Earth. One of the resonant states compatible with co-orbital behavior drives objects to a horseshoe 
         trajectory when plotted in a heliocentric frame of reference rotating with our planet because in this case the value of 
         $\lambda_{\rm r}$ oscillates about 180{\degr}, with an amplitude $>$240{\degr} \citep{1999ssd..book.....M}. Perturbed horseshoe 
         paths may lead to temporarily captured flybys, as in the case of 2022~NX$_{1}$ \citep{2022RNAAS...6..160D,2023A&A...670L..10D}, or 
         orbiters (see Appendix~\ref{tally}).

      \subsection{Data, data sources, and tools}
         Object A119q0V was initially reported on August 7, 2024, by the Asteroid Terrestrial-impact Last Alert System (ATLAS, 
         \citealt{2018PASP..130f4505T}) observing from Sutherland in South Africa. It was announced on August 14 with the provisional 
         designation 2024~PT$_{5}$ \citep{2024MPEC....P..170T}. Its orbital solution in Table~\ref{elements} is consistent with that of an
         Apollo-class asteroid. It is currently based on 358 observations with an observational timespan of 172~days and referred to epoch 
         JD 2460600.5 TDB, which is the origin of time in the calculations. It was retrieved from the Jet Propulsion Laboratory (JPL) 
         Small-Body Database (SBDB)\footnote{\href{https://ssd.jpl.nasa.gov/tools/sbdb\_lookup.html\#/}
         {https://ssd.jpl.nasa.gov/tools/sbdb\_lookup.html\#/}} provided by the Solar System Dynamics Group (SSDG, 
         \citealt{2011jsrs.conf...87G,2015IAUGA..2256293G}).\footnote{\href{https://ssd.jpl.nasa.gov/}{https://ssd.jpl.nasa.gov/}} The 
         object attracted our attention because it approached Earth at close range and low relative velocity.
%
%
      \begin{table}
       \centering
       \fontsize{8}{12pt}\selectfont
       \tabcolsep 0.14truecm
       \caption{\label{elements}Values of the heliocentric osculating orbital elements of 2024~PT$_{5}$ and their respective 1$\sigma$ 
                uncertainties.
               }
       \begin{tabular}{lcc}
        \hline
         Orbital parameter                                 &   & value$\pm$1$\sigma$ uncertainty \\
        \hline
         Semimajor axis, $a$ (au)                          & = &   1.012305060$\pm$0.000000009   \\
         Eccentricity, $e$                                 & = &   0.021476703$\pm$0.000000006   \\
         Inclination, $i$ (\degr)                          & = &   1.5205167$\pm$0.0000004       \\
         Longitude of the ascending node, $\Omega$ (\degr) & = & 305.572360$\pm$0.000007         \\
         Argument of perihelion, $\omega$ (\degr)          & = & 116.24844$\pm$0.00003           \\
         Mean anomaly, $M$ (\degr)                         & = & 323.67725$\pm$0.00003           \\
         Perihelion distance, $q$ (au)                     & = &   0.990564084$\pm$0.000000005   \\
         Aphelion distance, $Q$ (au)                       & = &   1.034046035$\pm$0.000000009   \\
         Absolute magnitude, $H$ (mag)                     & = &  27.4$\pm$0.5                   \\
        \hline
       \end{tabular}
       \tablefoot{The orbital solution of 2024~PT$_{5}$ is referred to epoch JD 2460600.5 (2024-Oct-17.0) TDB (Barycentric Dynamical 
                  Time, J2000.0 ecliptic and equinox), and  is based on 358 observations with an observational timespan of 172 days 
                  (solution date 2025-Jan-20 05:57:37 PST). Source: JPL's SBDB.
                 }
      \end{table}
%
%

         The orbit of 2024~PT$_{5}$ resembles that of 2022~NX$_{1}$, a confirmed natural object that experiences recurrent resonant 
         co-orbital episodes and temporarily captured flybys. Asteroid 2022~NX$_{1}$ also undergoes periodic close encounters with the 
         Earth--Moon  system that make the reconstruction of its past orbital evolution and the prediction of its future behavior beyond a 
         few decades difficult \citep{2022RNAAS...6..160D,2023A&A...670L..10D}. In such cases, the orbital evolution has to be studied 
         statistically considering the uncertainties of the orbit. The calculations needed to investigate the evolution of 2024~PT$_{5}$ 
         were carried out using a direct $N$-body code described by \citet{2003gnbs.book.....A},  publicly available from the website of 
         the Institute of Astronomy of the University of Cambridge,\footnote{\href{https://people.ast.cam.ac.uk/~sverre/web/pages/nbody.htm}
         {https://people.ast.cam.ac.uk/~sverre/web/pages/nbody.htm}} which implements the Hermite numerical integration scheme developed by 
         \citet{1991ApJ...369..200M}. Further details and the relevant results from this code were discussed in \citet{2012MNRAS.427..728D}. 
         Our physical model included the perturbations by the eight major planets, the Moon, the barycenter of the Pluto-Charon system, and 
         the 19 largest asteroids: Ceres, Pallas, Vesta, Hygiea, Euphrosyne, Interamnia, Davida, Herculina, Eunomia, Juno, Psyche, Europa, 
         Thisbe, Iris, Egeria, Diotima, Amphitrite, Sylvia, and Doris. For accurate initial positions and velocities (see 
         Appendix~\ref{Adata}), we used data based on the DE440/441 planetary ephemeris \citep{2021AJ....161..105P} from JPL's SSDG 
         {\tt Horizons} online Solar System data and ephemeris computation service.\footnote{\href{https://ssd.jpl.nasa.gov/horizons/}
         {https://ssd.jpl.nasa.gov/horizons/}} Most input data were retrieved from SBDB and {\tt Horizons} using tools provided by the 
         {\tt Python} package {\tt Astroquery} \citep{2019AJ....157...98G} and its {\tt HorizonsClass} 
         class.\footnote{\href{https://astroquery.readthedocs.io/en/latest/jplhorizons/jplhorizons.html}
         {https://astroquery.readthedocs.io/en/latest/jplhorizons/jplhorizons.html}}

   \section{Results\label{Results}}
      We used $N$-body simulations to assess the current dynamical status of 2024~PT$_{5}$, reflectance spectroscopy to determine its 
      physical nature, and photometry to study its rotational state.

      \subsection{Orbital evolution\label{ResultsO}}
\begin{figure*}
        \centering
         \includegraphics[width=\columnwidth]{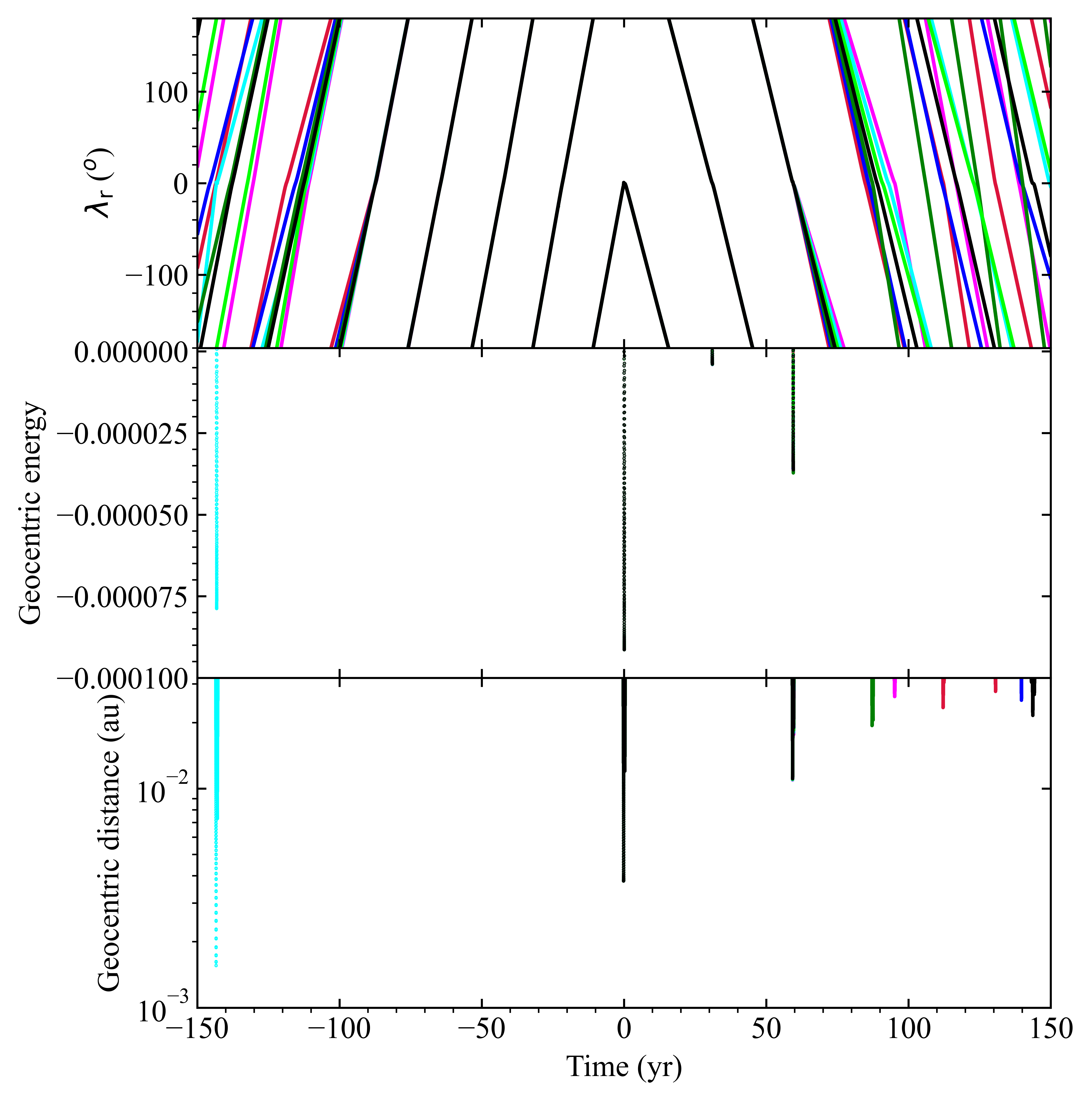}
         \includegraphics[width=\columnwidth]{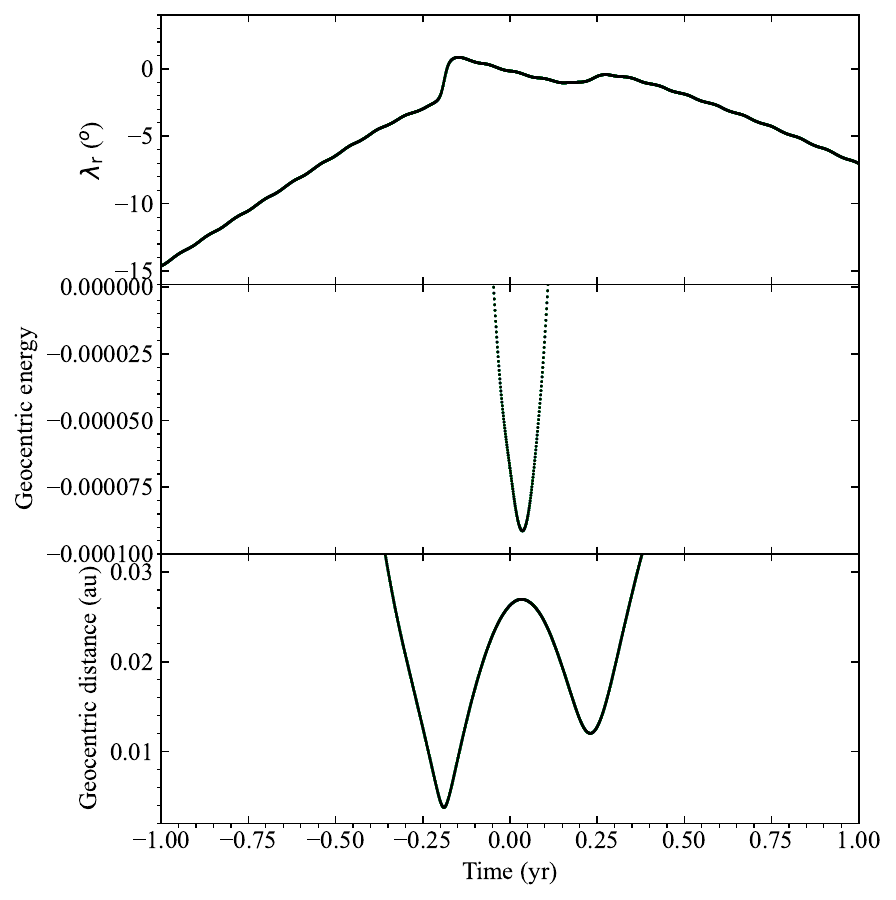}
         \caption{Short-term evolution and capture episode of 2024~PT$_{5}$. The left panels show the evolution of the values of the 
                  relevant parameters over the time interval (-150, 150)~yr. The right panels focus on the time window (-1.0, 1.0)~yr around 
                  the current epoch (2460600.5 TDB, 2024-Oct-17.0). Top: Time evolution of the relative mean longitude. Middle: Time 
                  evolution of the geocentric energy that focuses on negative values. Bottom: Time evolution of the geocentric distance for 
                  values under 0.03~au. The evolution of the nominal orbit is shown in black, those of the control orbits with state vectors 
                  separated $\pm3\sigma$ from the nominal ones in lime--green, $\pm6\sigma$ in cyan--blue, and $\pm9\sigma$ in 
                  fuchsia--crimson. The unit of energy is such that the units of mass, distance, and time are 1~$M_{\odot}$, 1~au, and one 
                  sidereal year divided by $2\pi$, respectively. The output interval is 0.36525~d.      
                 }
         \label{2024PT5ievolSIM}
      \end{figure*}

         Figure~\ref{2024PT5ievolSIM} summarizes the results of our calculations, showing the time evolution of the relevant parameters of 
         2024~PT$_{5}$: the relative mean longitude, and the geocentric energy and distance. The figure shows the evolution of the nominal 
         orbit and those of the control orbits or clones with state vectors (Cartesian coordinates and velocities; see 
         Table~\ref{vector2024PT5}) well away from the nominal orbit, up to $\pm9\sigma$ from the nominal orbital solution in 
         Table~\ref{elements}. The right panels in Fig.~\ref{2024PT5ievolSIM} show that the short-term evolution of all the control orbits 
         matches that of the nominal orbit. Therefore, we confirm the results in \citet{2024RNAAS...8..224D} obtained using an early orbital 
         solution. However, the left panel in Fig.~\ref{2024PT5ievolSIM} (see also Fig.~\ref{2024PT5ievolSIM123}) shows that recovering the 
         past orbital evolution of 2024~PT$_{5}$ prior to 1937 is difficult because a close encounter with Earth causes originally close 
         orbits to diverge. This is also found when predicting the behavior of this object beyond 2084. In other words, the orbital 
         evolution over the time interval ($-$87,~60)~yr in the left panel of Fig.~\ref{2024PT5ievolSIM} can be computed precisely, but 
         beyond this time interval the current orbital solution cannot provide reliable ephemerides.
%
%
%
%

         Figure~\ref{2024PT5ievolSIM}, right panels, shows that from 2023 until 2025, 2024~PT$_{5}$ followed a horseshoe-like trajectory
         (top panel) not too different from that of 2022~NX$_{1}$ in Fig.~3 of \citet{2023A&A...670L..10D}. It is also observed in 2024, 
         between September 29 and November 25, that the geocentric energy of all the control orbits had a negative value 
         (Fig.~\ref{2024PT5ievolSIM}, middle right panel), and that its geocentric distance remained under 0.03~au during the same time
         interval (bottom panel). According to \citet{1996Icar..121..207K}, 2024~PT$_{5}$ was a temporary satellite of Earth during that 
         period. Furthermore, as the asteroid failed to complete one revolution around Earth and following \citet{2012Icar..218..262G}, 
         2024~PT$_{5}$ underwent a temporarily captured flyby; we call this a quasi-mini-moon event. This result is statistically robust and 
         consistent across the control orbits tested (10$^{4}$). Figure~\ref{2024PT5ievolSIM}, left panels, shows that the evolution of 
         2024~PT$_{5}$ is seriously affected by orbital chaos induced by close encounters with the Earth--Moon system. The middle and bottom 
         panels show that recurrent temporarily captured flyby episodes could be possible. In fact, this object may experience further 
         temporarily captured flybys in the future (in 2084 for about 43~d).
 
      \subsection{Spectroscopy}
\begin{figure}
        \centering
         \includegraphics[width=\columnwidth]{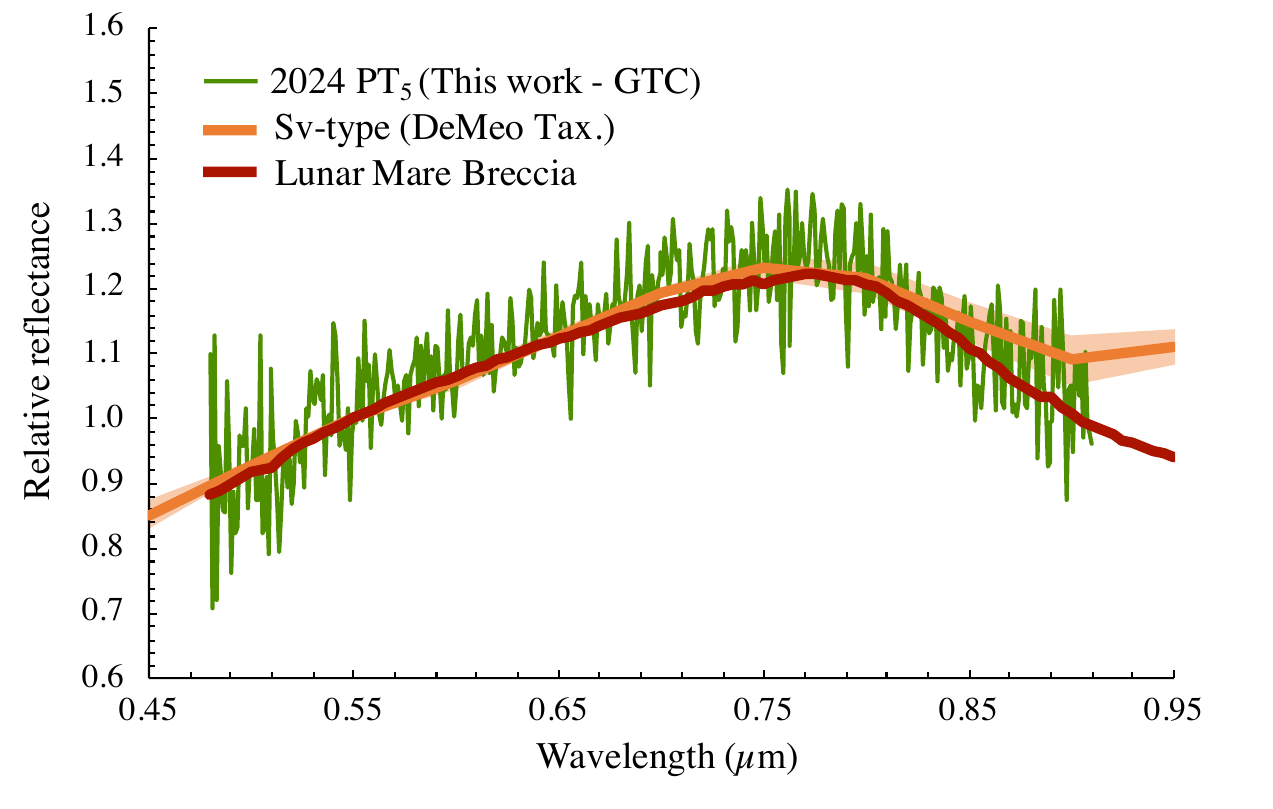}
         \caption{Visible spectrum of 2024~PT$_{5}$ obtained with GTC, in green, compared with the best taxonomical fit (Sv-type from 
                  \citealt{2009Icar..202..160D}), in orange. The hatched light orange area corresponds to $\pm$1$\sigma$ of the mean 
                  spectrum of Sv-types. A best-fit spectrum of a RELAB powder sample (ID LU-CMP-004-B) of a mare breccia from the Moon 
                  (Luna~24 mission) is also shown as a comparison, in red. The spectra were normalized to unity at 0.55~$\mu$m.
                  }
         \label{spectrum0}
      \end{figure}

         Three visible spectra of 2024~PT$_{5}$ were obtained on the night of September 7, 2024, starting at 22:45 UTC, using the OSIRIS 
         camera spectrograph \citep{2000SPIE.4008..623C,2010ASSP...14...15C}. The instrument is installed at the 10.4~m Gran Telescopio 
         Canarias (GTC), located at the El Roque de los Muchachos Observatory, on La Palma. It has a blue-sensitive 4096$\times$4096 pixel 
         CCD that yields a 7.8{\arcmin}$\times$7.8{\arcmin} field of view. Observations were done using a slit width of 1.2{\arcsec} 
         (oriented to the parallactic angle) and the R300R grism (0.48--0.92~$\mu$m). We obtained three individual spectra of 900~s of 
         exposure time each, at an airmass of 1.45. Two solar analog stars (Landolt SA 112-1333 and SA115-271) were also observed with the 
         same instrumental configuration and at a similar airmass to that of the asteroid in order to get its reflectance spectrum. The 2D 
         spectral images of the asteroid and the two solar analogs were bias and flat-field corrected, background subtracted, and collapsed 
         to 1D using an aperture equal to the distance from the center of the spatial profile to the pixel with 10\% peak intensity. 
         One-dimensional spectra were then wavelength calibrated with Xe+Ne+HgAr arc lamps, and a final averaged spectrum for the asteroid 
         and for each star were obtained. As a last step, we divided the spectrum of the asteroid by the spectrum of each solar analog star, 
         and averaged the two resulting ratios. The final reflectance spectrum of 2024~PT$_{5}$ is shown in Fig.~\ref{spectrum0}, in green.
%
%
%
%

         We used the M4AST\footnote{\href{https://spectre.imcce.fr/m4ast/index.php/index/home}
         {https://spectre.imcce.fr/m4ast/index.php/index/home}} online tool \citep{2012A&A...544A.130P} to get the most representative 
         taxonomical type for 2024~PT$_{5}$. This tool fits a curve to the spectrum and compares it to the ``taxons'' defined by 
         \citet{2009Icar..202..160D} using a $\chi^2$ procedure. The best fit corresponds to an Sv-type (in orange in 
         Fig.~\ref{spectrum0}), a class that serves as a link between S-types (mixtures of silicates and metal) and V-types, indicative of 
         basaltic surfaces, and with asteroid (4) Vesta as the main representative of the class. The M4AST tool was also used to compare the 
         spectrum of the asteroid with more than 15\,000 spectra of meteorites, terrestrial rocks, and lunar soils contained in the 
         RELAB\footnote{\href{https://sites.brown.edu/relab/relab-spectral-database/}
         {https://sites.brown.edu/relab/relab-spectral-database/}} database. Figure~\ref{spectrum0} shows the best fit in red, corresponding 
         to a powder sample (grain sizes from 10 to 45~$\mu$m) from a mare breccia of the Moon, collected by the Luna~24 mission. 
         Interestingly, and in order of increasing values of $\chi^2$, this best fit is followed by several samples from Moon maria 
         collected at the Apollo~12 landing site, and a sample of a terrestrial basaltic rock (Brockenheim, Germany).
%
%
      \begin{figure}
        \centering
         \includegraphics[width=\columnwidth]{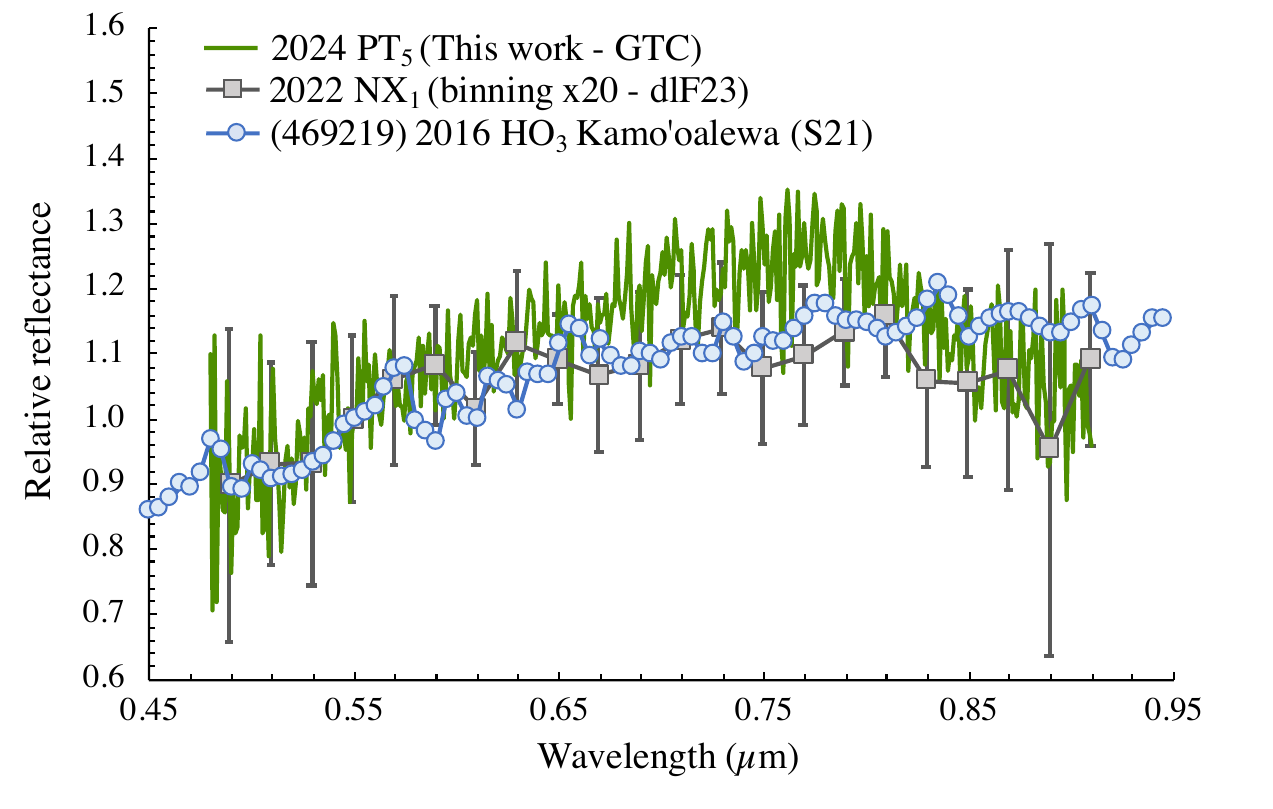}
         \caption{Comparison between the visible spectrum of 2022~NX$_{1}$, in gray (dlF23 -- \citealt{2023A&A...670L..10D}); 2024~PT$_{5}$, 
                  in green; and that of the quasi-satellite (469219) 2016~HO$_{3}$ Kamo'oalewa, in light blue (S21 -- 
                  \citealt{2021ComEE...2..231S}), all normalized to unity at 0.55~$\mu$m.
                  }
         \label{spectrum1}
      \end{figure}
%
%

         \citet{2021ComEE...2..231S} found that Earth's quasi-satellite (469219) 2016~HO$_{3}$ Kamo'oalewa \citep{2016MNRAS.462.3441D} has a 
         visible spectrum consistent with the S- and L-type taxonomic classes (see Fig.~\ref{spectrum1}), similar to that of 2022~NX$_{1}$ 
         published in \citet{2023A&A...670L..10D} and classified as a K-type, and different from the visible spectrum of 2024~PT$_{5}$. 
         However, when extended to the near-infrared, Kamo'oalewa shows a very red spectral slope, consistent with space-weathered 
         lunar-like silicates. This outlines the importance of having spectral information at near-infrared wavelengths, even in the form of 
         colors, as in \citet{2021ComEE...2..231S}, to better constrain the surface composition of asteroids. We note that our visible
         spectrum of 2024~PT$_{5}$ shows a deeper absorption band at 1~$\mu$m compared to that of Kamo'oalewa, suggesting less weathered 
         lunar-like silicates.

      \subsection{Light curve}
         The OSIRIS instrument was also used to study the rotational properties of the object. On September 28, 2024, a series of continuous 
         30~s exposures was taken over one hour using the Sloan-$r$ filter and the standard CCD configuration ($2\times2$ binning, plate 
         scale of 0.254~\arcsec~pixel$^{-1}$, readout time of 21~s). At the time of observation, the object had an apparent magnitude of 
         $r{\sim}23$ and a total apparent angular rate of 1.2~\arcsec~min$^{-1}$. Standard bias and flat-field corrections were applied to 
         reduce the images. 

\begin{figure}
        \centering
         \includegraphics[width=\columnwidth]{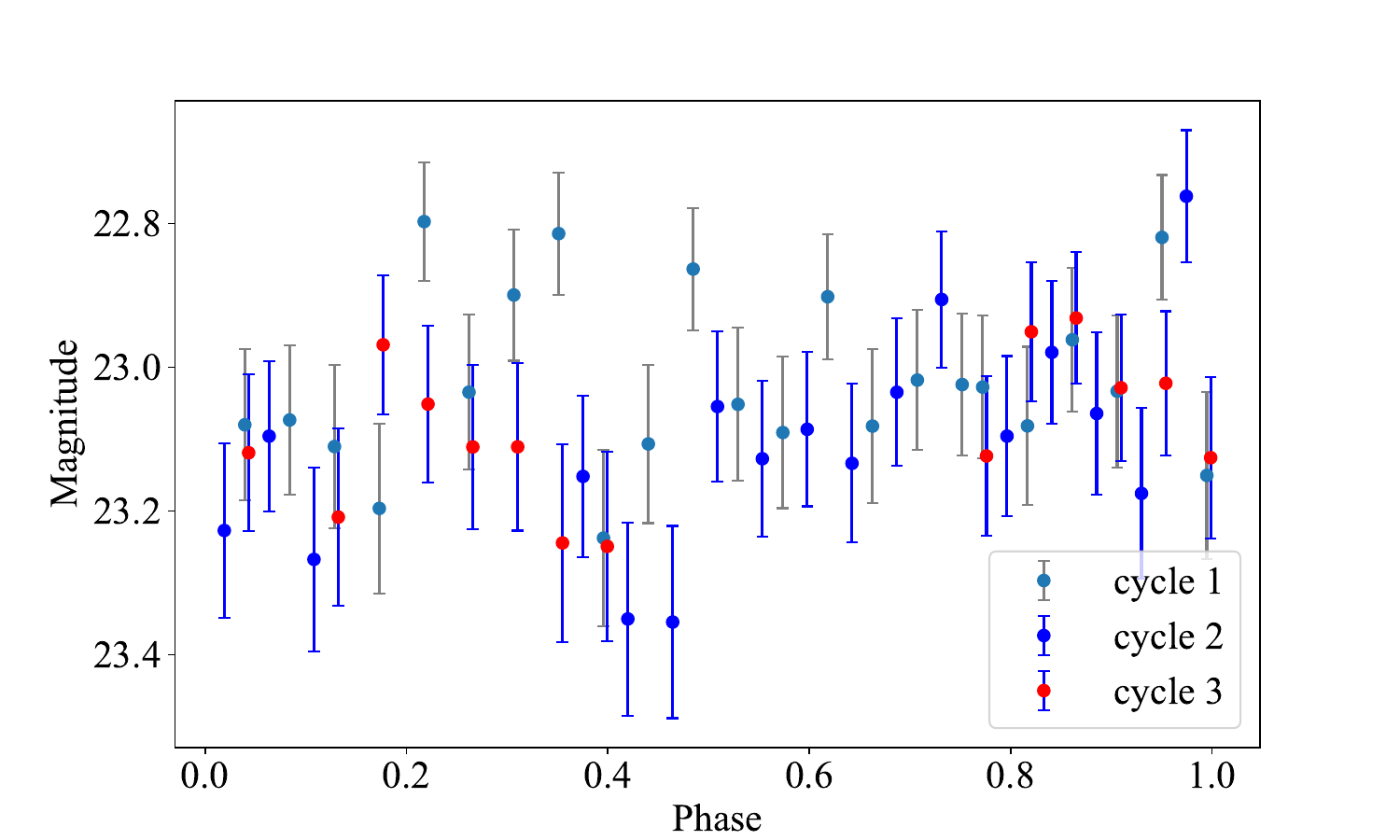}
         \caption{Light curve of 2024~PT$_{5}$ obtained with GTC/OSIRIS. The data are displayed binned in three cycles for a rotation period 
                  of 20.96~min. 
                  }
         \label{2024PT5LC}
      \end{figure}

         Relative aperture photometry was performed on the images using three field stars as reference stars, scaled by the mean magnitude 
         of the target during the observed interval. Different aperture radii were tested, including both fixed and variable apertures with 
         aperture corrections. The best result, obtained using a fixed aperture radius of about $1.25\times$FWHM, is shown in 
         Fig.~\ref{2024PT5LC}.

         The phased light curve is shown in Fig.~\ref{2024PT5LC} for a possible value of the period of 21$\pm$2~min (see further details in 
         Appendix~\ref{TSA}). A variability with an amplitude of $\sim$0.3--0.4~mag was observed during the one-hour observing block, which 
         is consistent with the uncertainty in $H$ in Table~\ref{elements}. We also note that the low signal-to-noise ratio (S/N) of the 
         data do not allow an unambiguous determination of the rotation period; but our data, together with those in 
         \citet{2025ApJ...978L..37B}, suggest a rotation period $\lesssim$1~h.  
%
%
%
%

   \section{Discussion\label{Discussion}}
      Our spectroscopic results on the surface composition of 2024~PT$_{5}$ are not conclusive due to the lack of a near-infrared 
      reflectance spectra, but they are suggestive of a lunar origin. \citet{2025ApJ...979L...8K} presented a visible to near-infrared 
      spectrum of 2024~PT$_{5}$ that, as in our case, is well matched by samples of the Moon, and is indicative of the presence of less 
      weathered lunar-like silicates for 2024~PT$_{5}$ than for (469219) 2016~HO$_{3}$ Kamo'oalewa and 2022~NX$_{1}$. Everything considered, 
      the three objects may all have a common origin as lunar ejecta. 

      The topic of the dynamics of lunar ejecta in the neighborhood of the Earth--Moon system was first studied by \citet{1994Icar..111..338W} 
      and \citet{1995Icar..118..302G}, who arrived at the conclusion that these materials can be scattered out of their source region relatively 
      quickly, on an effective timescale of $\sim$10$^5$~yr. This subject is now receiving renewed attention \citep{2023ComEE...4..372C,
      2025Icar..42916379C,2024NatAs...8..819J} after the confirmation of Kamo'oalewa as lunar ejecta, but it is also being linked to the 
      origin of the NEOs involved in temporary capture events \citep{2024AAS...24336303J}. Nevertheless, the theoretical expectations drawn 
      from the literature indicate that Kamo'oalewa, 2022~NX$_{1}$, and 2024~PT$_{5}$ may have all emerged out of lunar impact craters 
      formed during the last one million years or so. 

      Our rotation period results are consistent with the $\sim$20~min periodic variation reported by \citet{2025ApJ...978L..37B}. The 
      different results obtained by \citet{2025ApJ...979L...8K} can be attributed to changed viewing geometry. The obtained periodic 
      variability would be similar to that of Kamo'oalewa, which has an estimated rotation period of $28.3^{+1.8}_{-1.3}$~min 
      \citep{2021ComEE...2..231S}. For an object with an absolute magnitude of 27.4, the possibility of a higher rotational frequency or 
      even tumbling motion cannot be excluded. However, observations with higher time resolution and better S/N are required to confirm any 
      such behavior.

   \section{Summary and conclusions\label{Conclusions}}
      In this Letter we presented a detailed analysis of a mini-moon-like engagement experienced by Arjuna asteroid 2024~PT$_{5}$ in 2024, 
      from September 29 to November 25. The study of the short-term orbital evolution of 2024~PT$_{5}$ was carried out using direct $N$-body 
      simulations. We derived its spectral class from a visible reflectance spectrum obtained with the OSIRIS camera spectrograph at the 
      10.4~m Gran Telescopio Canarias. Data from the same instrument were used to perform a preliminary assessment of its rotational state. 
      Astrometry was obtained using the Two-meter Twin Telescope (see Appendix~\ref{sec:TTT}). Our conclusions can be summarized as follows:
      \begin{enumerate}
         \item We confirm that 2024~PT$_{5}$ is a natural object with a visible spectrum consistent with that of lunar mare breccia, 
               suggesting the presence of less weathered lunar-like silicates than in the case of (469219) 2016~HO$_{3}$ Kamo'oalewa or 
               2022~NX$_{1}$.
         \item We estimate a rotational period $\lesssim$1~h, tumbling not excluded.
         \item We confirm that 2024~PT$_{5}$ had negative geocentric orbital energy while inside a geocentric distance under three Hill 
               radii between September 29 and November 25, 2024. We call such engagements quasi-mini-moon events.
         \item The orbital evolutions of 2022~NX$_{1}$ and 2024~PT$_{5}$ as Arjunas are very similar.
      \end{enumerate}
      Both ESA and NASA now emphasize a low-cost approach to NEO missions for small body science and planetary defense, including the reuse 
      and repurposing of already active missions \citep{2024AcAau.224..122F}. The discovery and characterization of accessible objects such 
      as 2022~NX$_{1}$ and 2024~PT$_{5}$ using ground-based facilities is a necessary step before deciding whether an ongoing mission can be 
      extended to study a newly found NEO or if a new low-cost mission is perhaps preferable.

   \begin{acknowledgements}
      We thank the referee, M. Granvik, for a constructive, detailed, and actionable report. JdL and JL acknowledge support from the Spanish 
      `Agencia Estatal de Investigaci\'on del Ministerio de Ciencia e Innovaci\'on' (AEI-MCINN) under the grant PID2020-120464GB-I00. This 
      work was partially supported by the Spanish `Agencia Estatal de Investigaci\'on (Ministerio de Ciencia e Innovaci\'on)' under grant 
      PID2020-116726RB-I00 /AEI/10.13039/501100011033. Based on observations made with the Gran Telescopio Canarias (GTC), installed at the 
      Spanish Observatorio del Roque de los Muchachos of the Instituto de Astrof\'{\i}sica de Canarias, on the island of La Palma. This work 
      is partly based on data obtained with the instrument OSIRIS, built by a Consortium led by the Instituto de Astrof\'{\i}sica de 
      Canarias in collaboration with the Instituto de Astronom\'{\i}a of the Universidad Nacional Aut\'onoma de Mexico. OSIRIS was funded by 
      GRANTECAN and the National Plan of Astronomy and Astrophysics of the Spanish Government. This Letter includes observations made in the 
      Two-meter Twin Telescope (TTT) and Transient Survey Telescope (TST) in the Teide Observatory of the IAC, that Light Bridges operates 
      in the Island of Tenerife, Canary Islands (Spain). The Observing Time Rights (DTO) used for this research were provided by PEI project 
      SIDERA24. This Letter used flash storage and GPU computing resources as Indefeasible Computer Rights (ICRs) being commissioned at the 
      ASTRO POC project that Light Bridges will operate in the Island of Tenerife, Canary Islands (Spain). The ICRs used for this research 
      were provided by Light Bridges in cooperation with Hewlett Packard Enterprise (HPE) and VAST DAT. This research utilizes spectra 
      acquired by Carle M. Pieters with the NASA RELAB facility at Brown University. In preparation of this Letter, we made use of the NASA 
      Astrophysics Data System, the ASTRO-PH e-print server, and the Minor Planet Center (MPC) data server. 
   \end{acknowledgements}

   \bibliographystyle{aa}

   \begin{appendix}
      \section{One Hill radius versus three Hill radii\label{HillR}}
         A gravitationally bound two-body system has negative total energy. Although within the two-body problem this criterion is 
         unambiguous, subsystems within complex $N$-body systems may require additional constraints to ensure that a certain subsystem is
         indeed bound. In the particular case of planets and captured objects, \citet{1979RSAI...22..181C} advocated the use of a simple
         criterion, that the planetocentric energy of the object must be negative. \citet{1981A&A...102..165R} recommended adding, as a 
         second restriction, that the object completes at least one revolution around the planet. 

         \citet{1996Icar..121..207K} recognized the critical role played by the relative distance during temporary captures by planets 
         hosted by stars and defined temporary satellite capture by a planet using two conditions that must be met at the same time: The 
         planetocentric energy of the object must be negative and its planetocentric distance must be under three Hill radii for the planet 
         involved. The value of the Hill radius of Earth is roughly 0.01~au so the second condition for our planet implies flybys closer 
         than $\sim$0.03~au. This criterion involving both geocentric energy and separation was used by \citet{2012Icar..218..262G} to 
         define what temporarily captured natural irregular satellites of Earth are. From there, they designated captured objects that do 
         not complete at least one revolution around Earth as temporarily captured flybys and those completing one or more revolutions as 
         temporarily captured orbiters. However, \citet{2017Icar..285...83F} made the temporary capture criterion more restrictive by
         decreasing the separation threshold from three Hill radii to one Hill radius. For \citet{2017Icar..285...83F}, temporarily captured 
         flybys must have negative planetocentric orbital energy and maximum planetocentric distance under one Hill radius.

         In this work, we use the definitions in \citet{1996Icar..121..207K} and \citet{2012Icar..218..262G}. For a temporary capture to 
         take place, the intruding NEO must approach at close range ($<$0.03~au) and low relative velocity ($\lesssim$1~km~s$^{-1}$). For
         \citet{2017Icar..285...83F}, the unusual choreography interpreted by 2024~PT$_{5}$ and the Earth--Moon system is not a temporary
         capture.

         The Hill radius is often used to define the sphere of influence of a planet within the context of the two-body problem and the 
         restricted three-body problem \citep{1999ssd..book.....M}. As in \citet{2017Icar..285...83F}, it is often argued that true 
         temporary captures can only take place when the planetocentric energy becomes negative inside the Hill radius of the planet. We 
         believe that the restrictive use of the one Hill radius versus the three Hill radii criterion may lead to inconsistent 
         interpretations of the actual temporary orbital status of these objects. In fact, if calculations are repeated giving negligible 
         mass to the Earth--Moon system, no looping around Earth is observed; in other words, the gravitational influence of Earth in these 
         cases is not negligible beyond one Hill radius. In addition, the strict one revolution around Earth condition cited by some is 
         often difficult to assess because the geocentric loops traveled during temporary captures are far from closed elliptical orbits.

      \section{Mini-moons: The tally so far\label{tally}}
         So far, temporary captures by Earth, meeting the conditions in \citet{1996Icar..121..207K} and \citet{2012Icar..218..262G}, have 
         been reported for Apollo-class asteroids 2006~RH$_{120}$ \citep{2009A&A...495..967K}, 2020~CD$_{3}$ \citep{2020ApJ...900L..45B,
         2020MNRAS.494.1089D}, 2022~NX$_{1}$ \citep{2022RNAAS...6..160D,2023A&A...670L..10D}, and 2024~PT$_{5}$ \citep{2024RNAAS...8..224D,
         2025ApJ...978L..37B}. However, considering the more restrictive criterion featured in \citet{2017Icar..285...83F}, only 
         2006~RH$_{120}$, 2020~CD$_{3}$, and 2022~NX$_{1}$ experienced temporary captures by Earth. 

\begin{figure}
        \centering
         \includegraphics[width=\columnwidth]{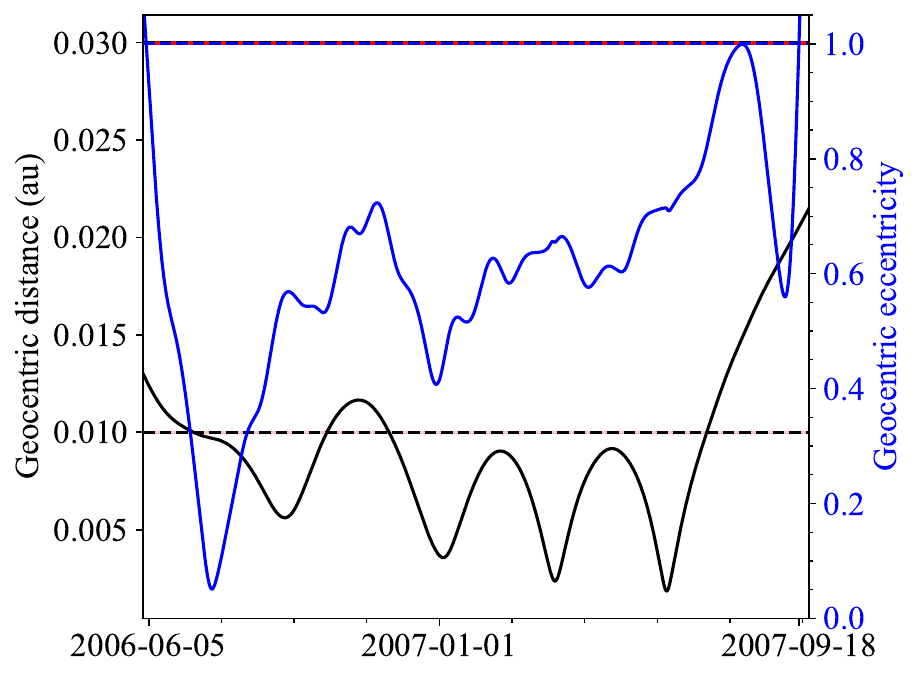}
         \includegraphics[width=\columnwidth]{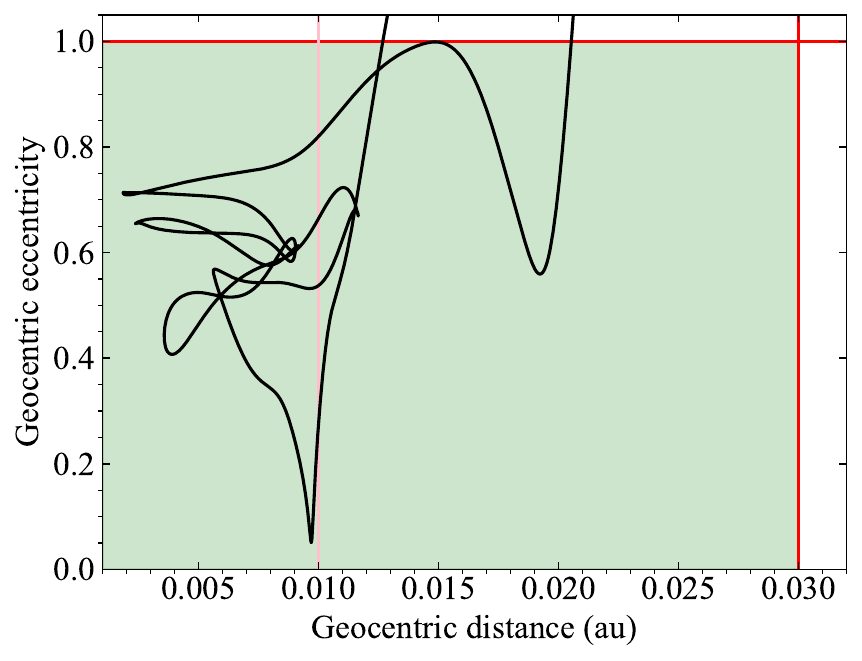}
         \caption{Capture episode for 2006~RH$_{120}$.
                  {\it Top panel:} Time evolution of the geocentric eccentricity and distance. 
                  {\it Bottom panel:} Geocentric eccentricity as a function of the geocentric distance. The one Hill radius mark is shown in 
                  pink, the three Hill radii mark in red. Source: JPL's {\tt Horizons}.
                 }
         \label{2006RH120JPL}
      \end{figure}

\begin{figure}
        \centering
         \includegraphics[width=\columnwidth]{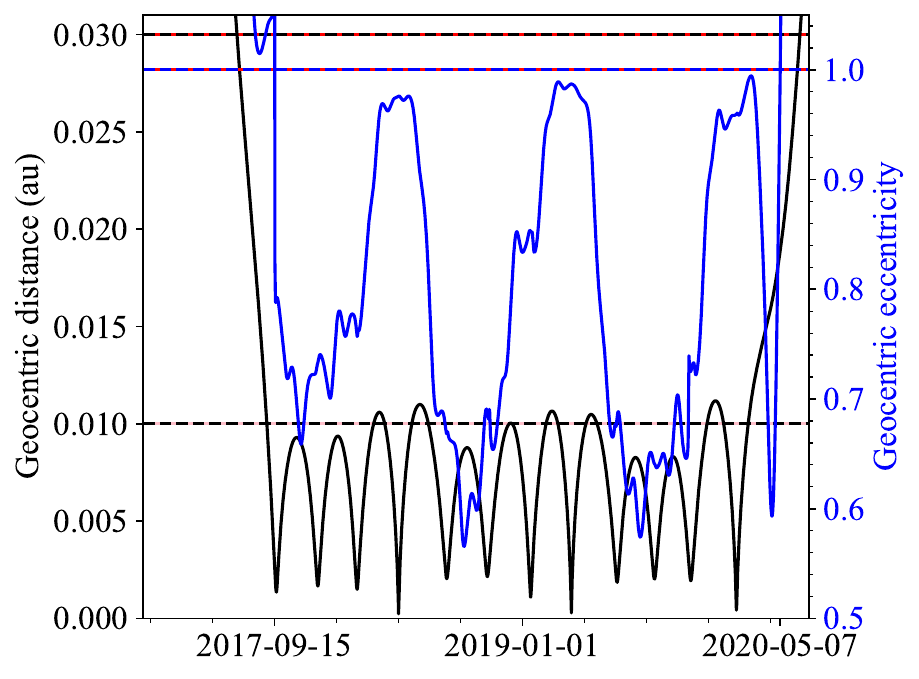}
         \includegraphics[width=\columnwidth]{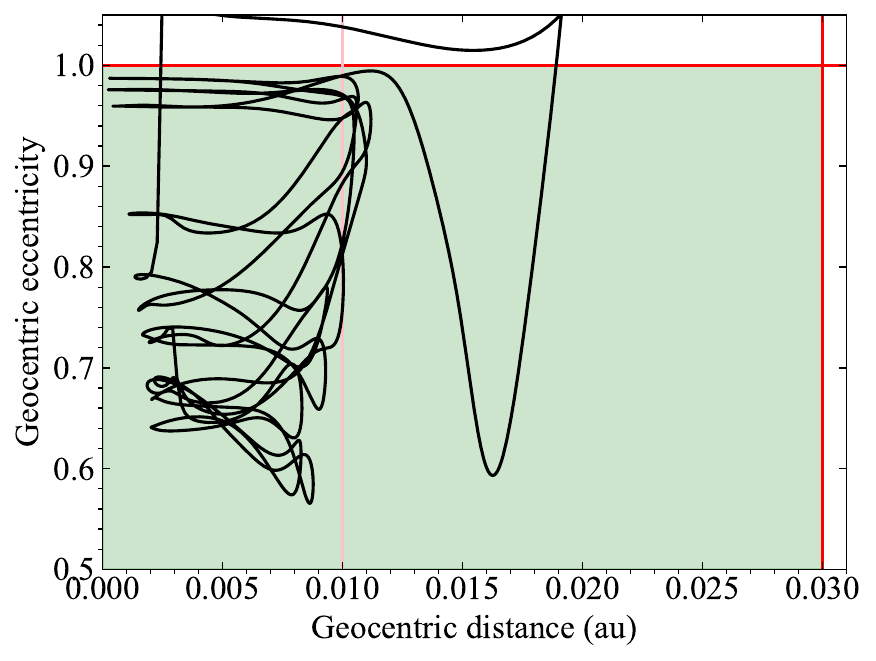}
         \caption{Capture episode for 2020~CD$_{3}$.
                  {\it Top panel:} Time evolution of the geocentric eccentricity and distance. 
                  {\it Bottom panel:} Geocentric eccentricity as a function of the geocentric distance. The one Hill radius mark is shown in 
                  pink, the three Hill radii mark in red. Source: JPL's {\tt Horizons}.
                 }
         \label{2020CD3JPL}
      \end{figure}

\begin{figure}
        \centering
         \includegraphics[width=\columnwidth]{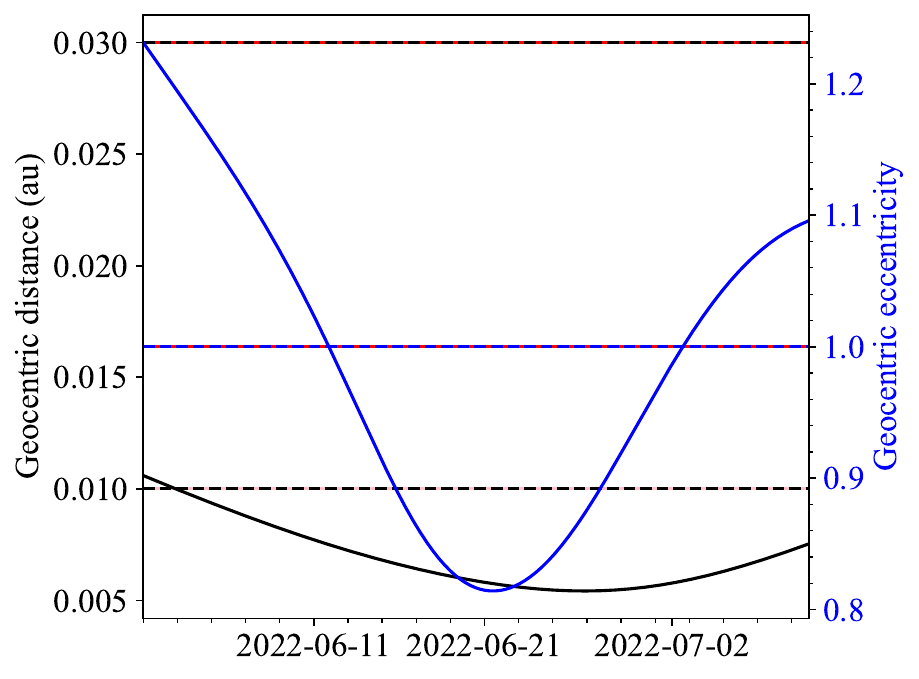}
         \includegraphics[width=\columnwidth]{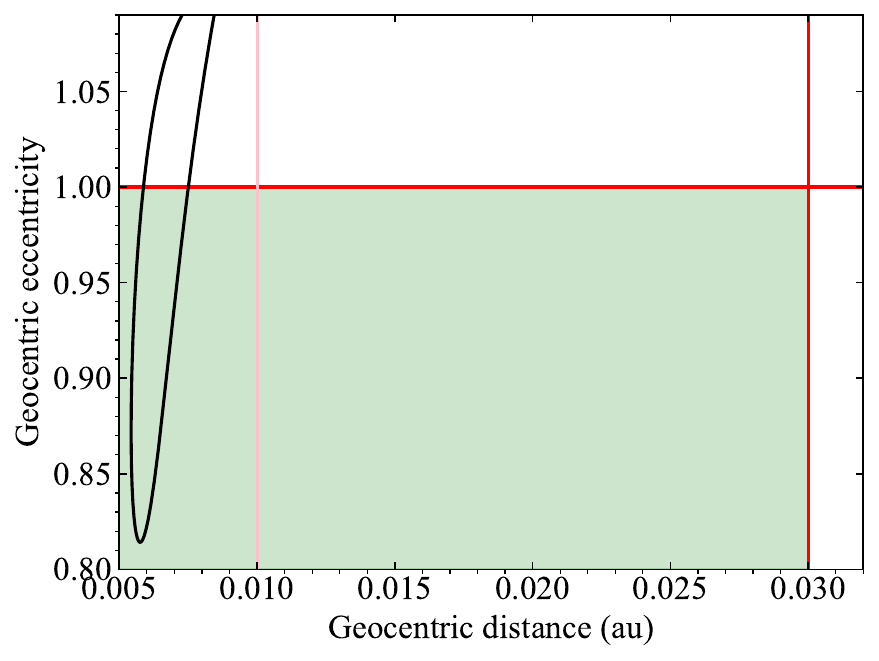}
         \caption{Capture episode for 2022~NX$_{1}$.
                  {\it Top panel:} Time evolution of the geocentric eccentricity and distance. 
                  {\it Bottom panel:} Geocentric eccentricity as a function of the geocentric distance. The one Hill radius mark is shown in 
                  pink, the three Hill radii mark in red. Source: JPL's {\tt Horizons}.
                 }
         \label{2022NX1JPL}
      \end{figure}

\begin{figure}
        \centering
         \includegraphics[width=\columnwidth]{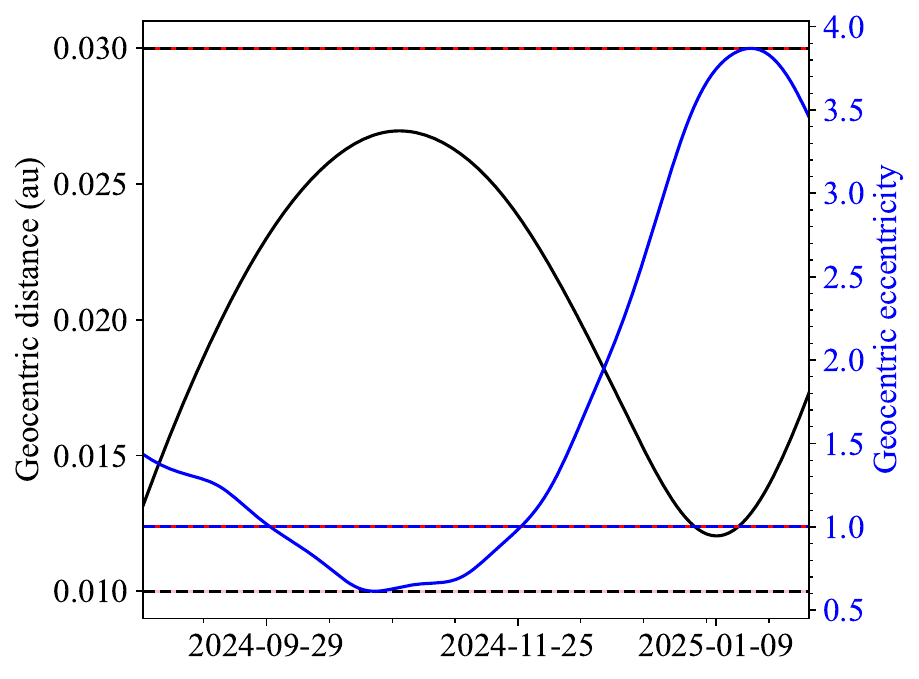}
         \includegraphics[width=\columnwidth]{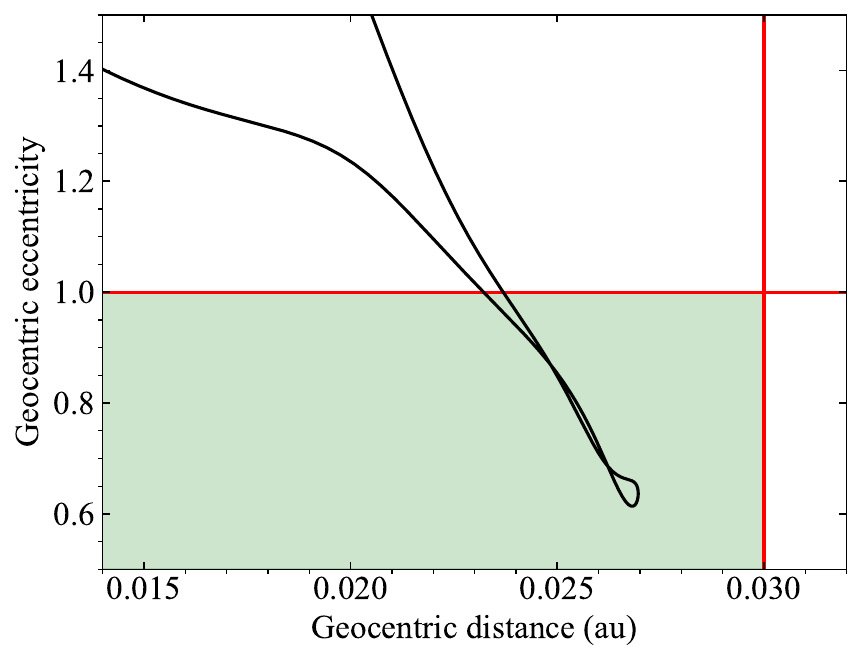}
         \caption{Capture episode for 2024~PT$_{5}$.
                  {\it Top panel:} Time evolution of the geocentric eccentricity and distance. 
                  {\it Bottom panel:} Geocentric eccentricity as a function of the geocentric distance. Source: JPL's {\tt Horizons}.
                 }
         \label{2024PT5JPL}
      \end{figure}

         The condition requiring negative planetocentric energy during capture is equivalent to having planetocentric eccentricity $<$1. 
         Figures~\ref{2006RH120JPL} to \ref{2024PT5JPL} are similar to Fig.~\ref{2024PT5SIM}, but they show the ephemerides retrieved from 
         JPL's {\tt Horizons} instead of displaying the results of our computer simulations. They clearly show the differences between 
         temporarily  captured orbiters (2006~RH$_{120}$ and 2020~CD$_{3}$) and temporarily captured flybys (2022~NX$_{1}$ and 
         2024~PT$_{5}$), but they also show that the strict application of the one Hill radius criterion may lead us to reclassify the 
         capture episodes of 2006~RH$_{120}$ and 2020~CD$_{3}$ as a sequence of temporarily captured flybys, which is obviously incorrect.
%
%
%
%
%
%
%
%
%
%
%
%
%
%
%
%

      \section{Input data and uncertainties\label{Adata}}
         Here we include the barycentric Cartesian state vector of NEO 2024~PT$_{5}$. This vector and its uncertainties were used to 
         perform the calculations discussed above and to generate the figures that display the time evolution of the critical angle, 
         $\lambda_{\rm r}$, and the geocentric energy and distance. As an example, a new value of the $X$-component of the state vector was 
         computed using $X_{\rm c} = X + \sigma_X \ r$, where $r$ is an univariate Gaussian random number, and $X$ and $\sigma_X$ are the 
         mean value and its 1$\sigma$ uncertainty in Table~\ref{vector2024PT5}.
%
%
     \begin{table}
      \centering
      \fontsize{8}{12pt}\selectfont
      \tabcolsep 0.15truecm
      \caption{\label{vector2024PT5}Barycentric Cartesian state vector of 2024~PT$_{5}$: Components and associated 1$\sigma$ uncertainties.
              }
      \begin{tabular}{ccc}
       \hline
        Component                         &   &    value$\pm$1$\sigma$ uncertainty                                 \\
       \hline
        $X$ (au)                          & = &    9.023402218777196$\times10^{-1}$$\pm$4.92897003$\times10^{-9}$  \\
        $Y$ (au)                          & = &    4.000187059808887$\times10^{-1}$$\pm$8.32396963$\times10^{-9}$  \\
        $Z$ (au)                          & = &    2.605307727298394$\times10^{-2}$$\pm$5.84709025$\times10^{-9}$  \\
        $V_X$ (au/d)                      & = & $-$7.275854648367192$\times10^{-3}$$\pm$9.95845747$\times10^{-11}$ \\
        $V_Y$ (au/d)                      & = &    1.578995907831493$\times10^{-2}$$\pm$6.93400853$\times10^{-11}$ \\
        $V_Z$ (au/d)                      & = &    8.655418453965767$\times10^{-5}$$\pm$7.24496957$\times10^{-11}$ \\
       \hline
      \end{tabular}
      \tablefoot{Data are referred to epoch JD 2460600.5, which corresponds to 0:00 on 2024 October 17 TDB (J2000.0 ecliptic and equinox). 
                 Source: JPL's {\tt Horizons}.
                }
     \end{table}
%
%

\begin{figure}
        \centering
         \includegraphics[width=\columnwidth]{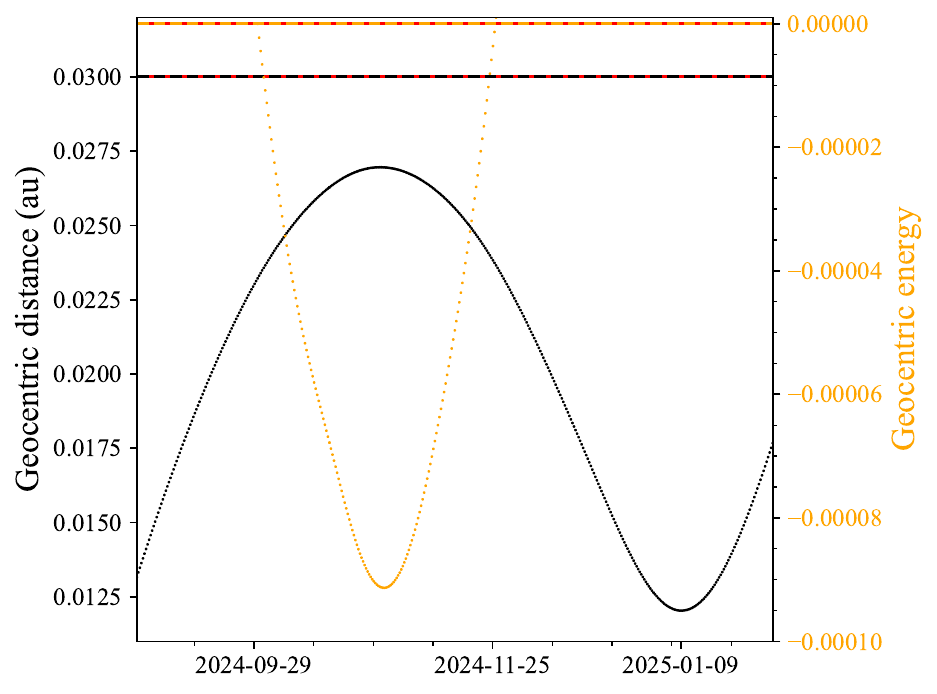}
         \includegraphics[width=\columnwidth]{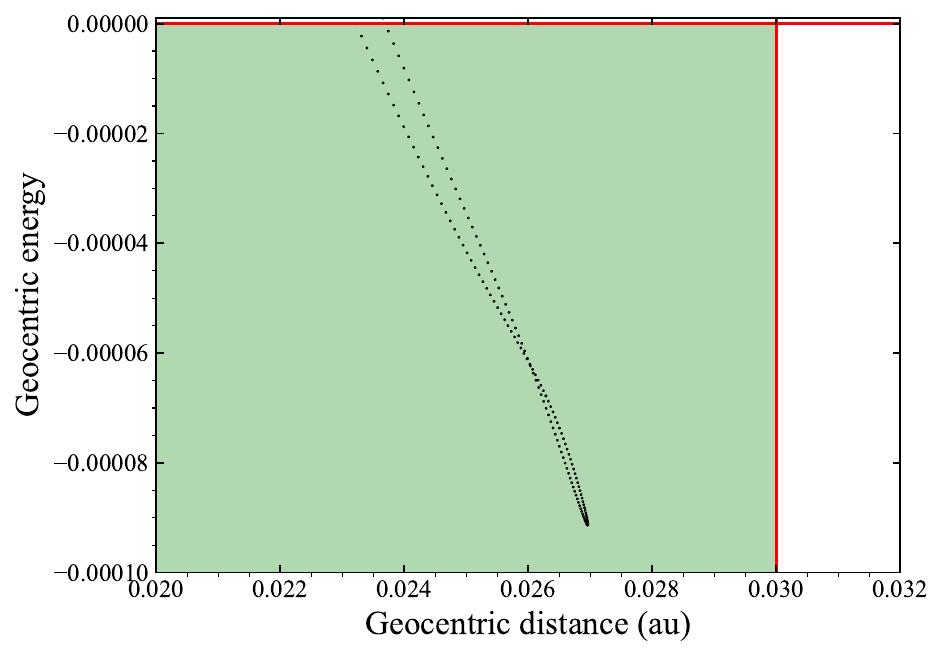}
         \caption{Capture episode for 2024~PT$_{5}$. {\it Top panel:} Time evolution of the geocentric energy and distance. {\it Bottom 
                  panel:} Geocentric energy as a function of the geocentric distance. The green zone signals the region where the definition 
                  in \citet{1996Icar..121..207K} holds. The unit of energy is such that the unit of mass is 1~$M_{\odot}$, the unit of 
                  distance is 1~au, and the unit of time is one sidereal year divided by $2\pi$.
                 }
         \label{2024PT5SIM}
      \end{figure}

         Figure~\ref{2024PT5SIM} shows the capture episode for the nominal orbit in greater detail. The minimum value of the geocentric 
         energy coincides with a local maximum in the geocentric distance. The capture takes place when the asteroid approaches the 
         apogee of its highly asymmetric geocentric trajectory and its geocentric velocity reaches a minimum.         
%
%
%
%

\begin{figure}
        \centering
         \includegraphics[width=\columnwidth]{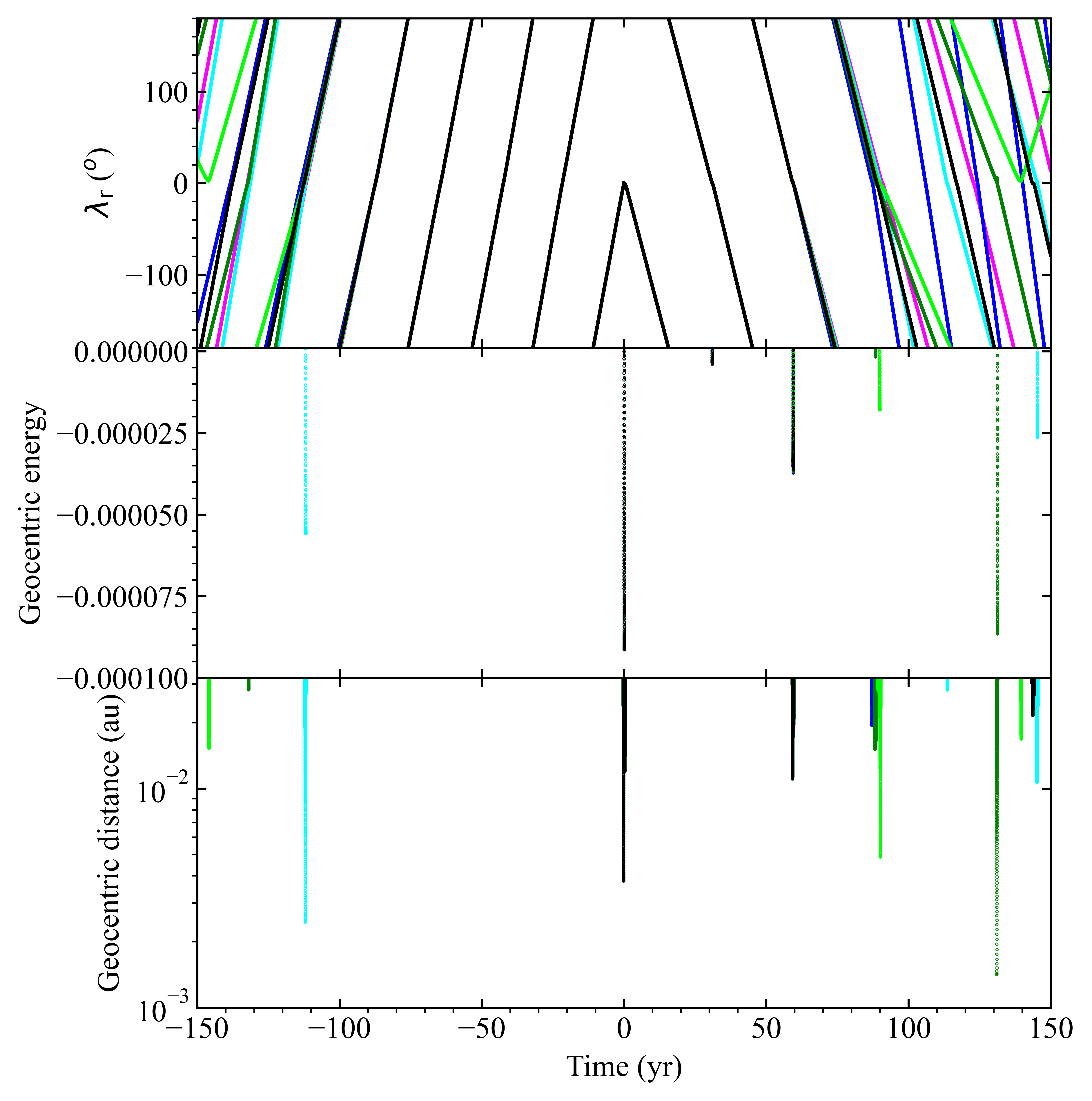}
         \caption{Short-term evolution and capture episode of 2024~PT$_{5}$. The panels show the evolution of the values of relevant 
                  parameters over the time interval ($-$150,~150)~yr around the current epoch. {\it Top panel:} Time evolution of the 
                  relative mean longitude. {\it Middle panel:} Time evolution of the geocentric energy that focuses on negative values. 
                  {\it Bottom panel:} Time evolution of the geocentric distance that focuses on values under 0.03~au. The evolution of the 
                  nominal orbit is shown in black, those of control orbits with state vectors separated $\pm1\sigma$ from the nominal ones 
                  in lime--green, $\pm2\sigma$ in cyan--blue, and $\pm3\sigma$ in fuchsia--crimson. The unit of energy is such that the unit 
                  of mass is 1~$M_{\odot}$, the unit of distance is 1~au, and the unit of time is one sidereal year divided by $2\pi$. The 
                  output interval is 0.36525~d. The origin of time is epoch 2460600.5 (2024-Oct-17.0) TDB.     
                 }
         \label{2024PT5ievolSIM123}
      \end{figure}

         Figure~\ref{2024PT5ievolSIM123} is analogous to Fig.~\ref{2024PT5ievolSIM}, but corresponds to control orbits with state vectors 
         separated by $\pm1\sigma$, $\pm2\sigma$, and $\pm3\sigma$ from that of the nominal orbit. These results are consistent with those 
         presented in Sect.~\ref{ResultsO}, the orbital evolution of 2024~PT$_{5}$ over the time interval ($-$87,~60)~yr is robust, but 
         outside this interval the dynamics have to be discussed in statistical terms as the evolution of arbitrarily close orbits diverges 
         exponentially. Figure~\ref{2024PT5ievolSIM123}, top panel, shows that 2024~PT$_{5}$ might have experienced 1:1 resonant behavior of 
         the horseshoe type in the past, and perhaps it will  again in the future, but the evolution is strongly affected by orbital chaos 
         driven by close encounters with the Earth--Moon system. 
%
%
%
%

      \section{Time series analysis\label{TSA}}
\begin{figure}
        \centering
         \includegraphics[width=\columnwidth]{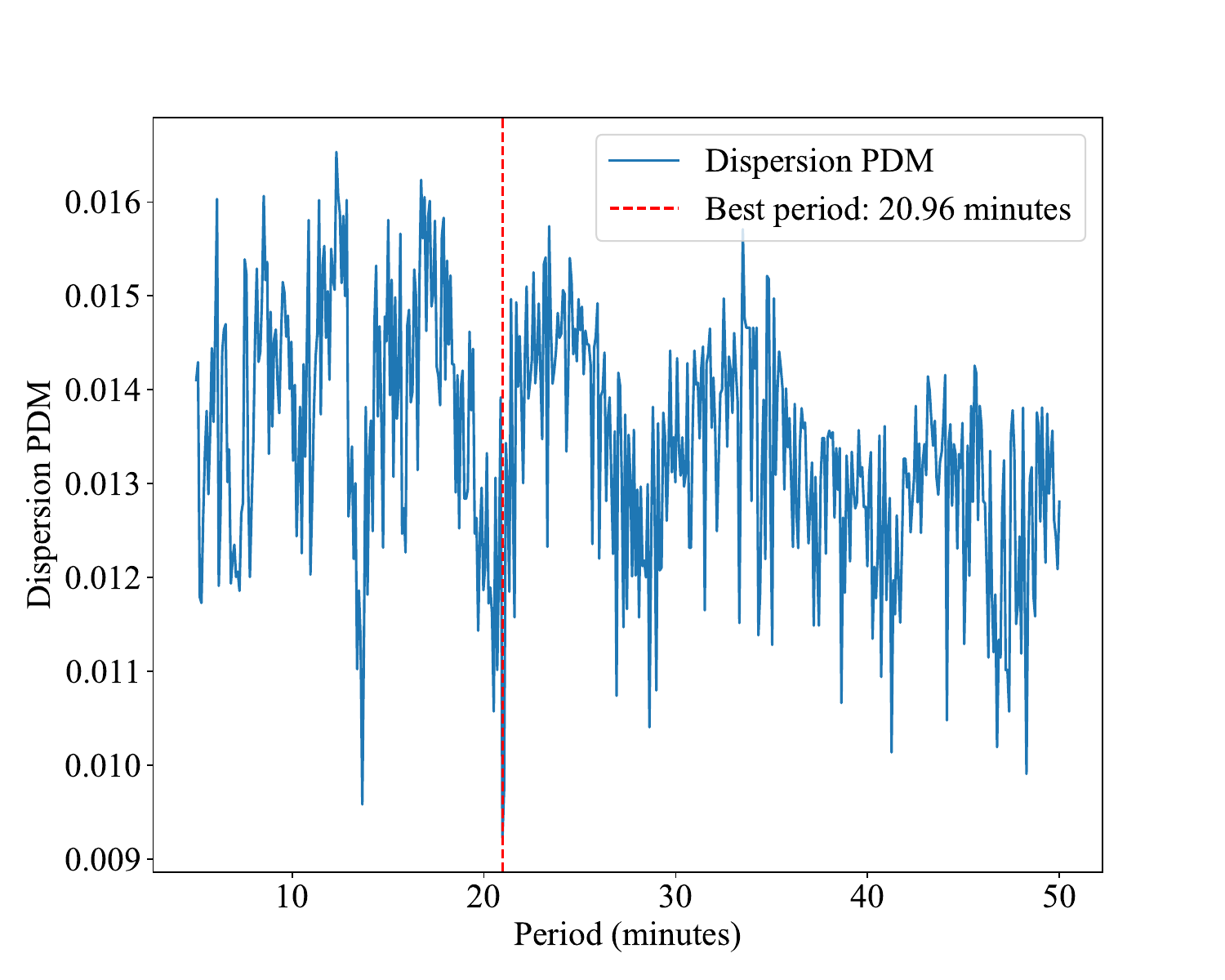}
         \caption{Sum of the squares of the differences in the ordinate from one data point to the next for the light curve of 2024~PT$_{5}$ 
                  obtained with GTC/OSIRIS. The value that produces the least possible scatter is $\sim$21~min. 
                  }
         \label{2024PT5D}
      \end{figure}

         We analyzed the light curve using the Phase Dispersion Minimization method (PDM; \citealt{1978ApJ...224..953S}) to find its
         rotation period. The PDM method minimizes the sum of the squares of the differences in the ordinate from one data point to the 
         next. The chosen period is the one resulting in the smallest sum; in other words, the one producing the least possible scatter 
         about the derived light curve. The dispersion as a function of the test period is shown in Fig.~\ref{2024PT5D}. We obtained a 
         possible variation with a period $P$=21$\pm$2~min.
%
%
%
%

      \section{TTT and TST astrometry \label{sec:TTT}}
\begin{figure}
        \centering
         \includegraphics[width=\columnwidth]{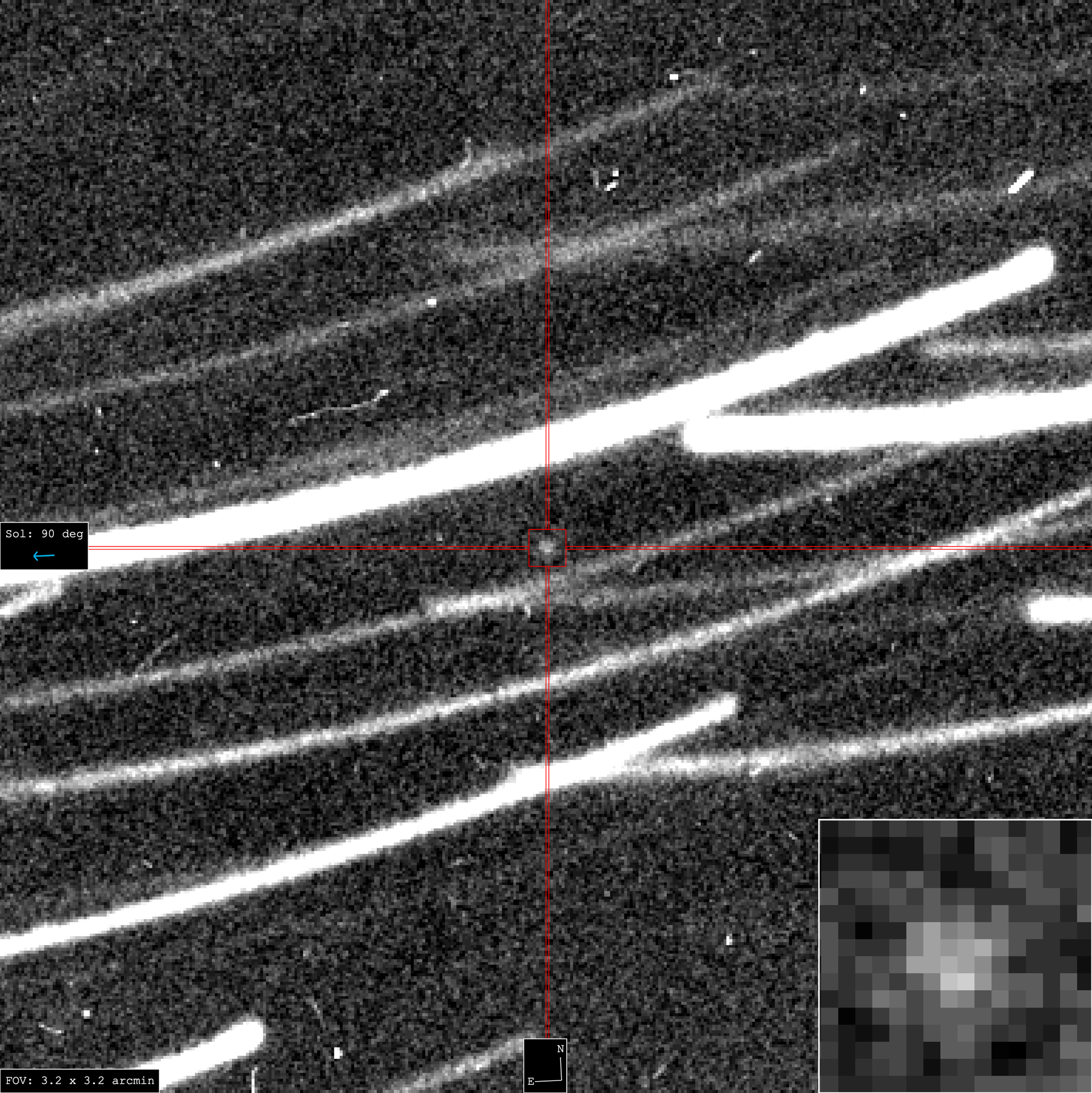}
         \caption{Image of 2024~PT$_{5}$ (center, point source) obtained with TTT1. This image is a combination of a sequence of individual
                  images, each  obtained in sidereal tracking mode. The asteroid remains stationary in the combined image because it was 
                  tracked and stacked using 2024~PT$_{5}$ ephemeris data from the MPC website (see  text for details). The background 
                  stars trail because of 2024~PT$_{5}$'s motion. The bottom right square shows a magnified view of the point-like source.
                 }
         \label{2024PT5TTT}
      \end{figure}

         Astrometric observations were obtained using three telescopes located at Teide Observatory (OT, Tenerife, Canary Islands, Spain), 
         the TTT1 and TTT2 (Two-meter Twin Telescope), and the TST (Transient Survey Telescope) as part of the key project observations 
         during the commissioning phase. The observational circumstances are shown in Table \ref{tab_obs_cir}. Additional astrometry from
         these telescopes was submitted to the Minor Planet Center 
         (MPC).\footnote{\url{https://minorplanetcenter.net/db_search/show_object?utf8=\%E2\%9C\%93&object_id=2024+PT5}} A stacked image of 
         2024~PT$_{5}$ obtained with TTT1 is shown in Fig.~\ref{2024PT5TTT}.
%
%
%
%

         TTT\footnote{\url{https://ttt.iac.es/}} currently has two telescopes, TTT1 and TTT2. They are two Ritchey-Chr\'etien optical 
         systems, with an aperture of 0.80\,m, altazimuthal mount, and two f/6.85 Nasmyth foci each. TST\footnote{\url{https://tst.iac.es/}} 
         is an ultra-wide-field prime corrector optical system, with an aperture of 1~m, equatorial mount, and a f/1.3 primary foci.

         TTT1 was equipped with a 2Kpx$\times$2Kpx Andor iKon-L 936 camera, with a back-illuminated 13.5~$\mu$m pixel$^{-1}$ BEX2-DD CCD 
         sensor, resulting in a FoV of 17.3{\arcmin}$\times$17.3{\arcmin}, and a plate scale of 0.51{\arcsec}~pixel$^{-1}$.
         QHY411M\footnote{\url{https://www.qhyccd.com/}} cameras \citep{2023PASP..135e5001A} are installed on TTT2 and TST telescopes. The 
         QHY411M camera sports a scientific Complementary Metal-Oxide-Semiconductor (sCMOS) image sensor with 14304~px~$\times$~10748~px 
         (151~Mpx~=~302~MB) of size 3.76~$\mu$m~pixel$^{-1}$. This setup provides an effective FoV of 34$^{\prime}\times$25$^{\prime}$ (with 
         a plate scale of 0.14{\arcsec}~pixel$^{-1}$) for TTT2 and 2.4{\degr}$\times$1.8{\degr} (0.6{\arcsec}~pixel$^{-1}$) for TST.

         All the images were taken using the $Luminance$  filter (Lum), that covers the 0.4 to 0.7~$\mu$m wavelength range. Data reduction 
         was done using standard procedures, correcting for bias, dark, and sky flat-fielding. The astrometry was extracted using the Tycho 
         software,\footnote{\url{https://www.tycho-tracker.com}} applying the track and stack method. To increase the object's S/N, stacking 
         was done tracking on 2024~PT$_{5}$ orbit using the orbital ephemeris of the object (individual images have sidereal tracking). The 
         ephemerides were obtained from the MPC website.\footnote{\url{https://www.minorplanetcenter.net/iau/MPEph/MPEph.html}}

%
%
      \begin{table*}
        \caption{Observing log.\label{tab_obs_cir}} 
        \centering                          
        \begin{tabular}{c c c c c c c}        
          \hline\hline                 
           Date obs.  & $UT_{start}$ & Obs & exp            & filter & Mag   & Telescope   \\    
          \hline                        
           2024-09-10 &  20:11       &   7 & 130$\times$60s & Lum    & 22.40 & Ikon-TTT1   \\
           2024-09-10 &  20:10       &   7 & 168$\times$60s & Lum    & 22.17 & QHY411-TTT2 \\
           2024-09-10 &  21:22       &  11 & 123$\times$60s & Lum    & 22.05 & QHY411-TST  \\
           2024-09-11 &  20:09       &   6 & 130$\times$60s & Lum    & 21.95 & Ikon-TTT1   \\
           2024-09-11 &  21:03       &   4 & 141$\times$60s & Lum    & 22.31 & QHY411-TST  \\
           2024-09-12 &  20:06       &   8 & 130$\times$60s & Lum    & 22.26 & Ikon-TTT1   \\
           2024-09-13 &  20:08       &   3 &  78$\times$60s & Lum    & 21.63 & Ikon-TTT1   \\
           2024-09-15 &  20:02       &   3 &  78$\times$60s & Lum    & 22.51 & Ikon-TTT1   \\
           2024-09-17 &  20:01       &   2 &  48$\times$60s & Lum    & 22.51 & Ikon-TTT1   \\
           2024-09-20 &  20:30       &   3 &  60$\times$60s & Lum    & 22.72 & Ikon-TTT1   \\
           2024-09-21 &  19:59       &   3 &  79$\times$60s & Lum    & 22.77 & Ikon-TTT1   \\
           2024-09-22 &  19:54       &   2 &  53$\times$60s & Lum    & 22.18 & Ikon-TTT1   \\
           2024-09-23 &  19:52       &   2 &  78$\times$60s & Lum    & 23.09 & Ikon-TTT1   \\
           2024-09-24 &  19:55       &   3 &  65$\times$60s & Lum    & 22.53 & Ikon-TTT1   \\
           2024-09-25 &  19:51       &   3 &  65$\times$60s & Lum    & 22.09 & Ikon-TTT1   \\
           2024-09-26 &  19:51       &   6 & 143$\times$60s & Lum    & 22.17 & Ikon-TTT1   \\
           2024-09-27 &  19:47       &   3 &  78$\times$60s & Lum    & 22.43 & Ikon-TTT1   \\
           2024-09-27 &  19:47       &   3 & 126$\times$60s & Lum    & 22.91 & QHY411-TTT2 \\
           2024-09-27 &  20:00       &   4 & 245$\times$30s & Lum    & 22.66 & QHY411-TST  \\
           2024-09-28 &  19:40       &   2 &  78$\times$60s & Lum    & 22.23 & Ikon-TTT1   \\
           2024-09-28 &  20:02       &   4 &  84$\times$90s & Lum    & 22.66 & QHY411-TST  \\
           2024-09-28 &  21:24       &   6 &  60$\times$30s & $r$    & 22.66 & GTC         \\
           2024-09-29 &  19:40       &   3 &  78$\times$60s & Lum    & 22.8  & Ikon-TTT1   \\
           2024-09-30 &  19:37       &   3 &  73$\times$60s & Lum    & 22.1  & Ikon-TTT1   \\
           2024-10-01 &  19:50       &   3 &  87$\times$60s & Lum    & 22.2  & Ikon-TTT1   \\
           2024-10-04 &  19:35       &   3 & 182$\times$60s & Lum    & 22.4  & Ikon-TTT1   \\
           2024-10-05 &  19:55       &   3 & 130$\times$60s & Lum    & 22.23 & Ikon-TTT1   \\
           2024-10-07 &  21:03       &   2 &  26$\times$60s & Lum    & 22.10 & Ikon-TTT1   \\
           2024-10-08 &  19:40       &   3 &  71$\times$60s & Lum    & 22.47 & Ikon-TTT1   \\
           2024-10-09 &  19:51       &   3 & 143$\times$60s & Lum    & 22.55 & Ikon-TTT1   \\
           2024-10-10 &  19:50       &   3 & 141$\times$60s & Lum    & 21.50 & Ikon-TTT1   \\
           2024-10-14 &  19:42       &   2 & 130$\times$60s & Lum    & 22.55 & Ikon-TTT1   \\
           2024-10-18 &  19:45       &   2 & 104$\times$60s & Lum    & 22.1  & Ikon-TTT1   \\
           2024-10-19 &  19:45       &   2 &  91$\times$60s & Lum    & 22.5  & Ikon-TTT1   \\
          \hline                                   
        \end{tabular}
        \tablefoot{The table includes the acquisition date of the sequence (Date), the starting time for the sequence ($UT_{start}$), the 
                   number of reported observations (Obs), the total exposure time expressed in seconds, the filter, the V magnitude derived 
                   from the stacked image, and the instrument (Telescope). Magnitudes give the V magnitude calculated from aperture 
                   photometry and were standardized using catalog sources. Due to low S/N of the asteroid and star-trail contamination, 
                   magnitude errors could be as high as 0.5~mag.}
      \end{table*}
%
%

   \end{appendix}

\end{document}